\documentclass[12pt]{article}
\usepackage[latin1]{inputenc}
\usepackage{amsfonts,amssymb,amsmath, epsfig}

\def\Box{\leavevmode\vbox{\hrule
     \hbox{\vrule\kern4pt\vbox{\kern4pt}%
           \vrule}\hrule}}
\def\blackbox{\leavevmode\vrule height 5pt width 4pt depth 0pt\relax}
\def\endproof{\null\hfill {$\blackbox$}\bigskip}

\newcounter{appendix}
\def\appendix{\advance\c@appendix by 1
   \def\thesection{\Alph{section}}
   \ifnum\c@appendix=1 \setcounter{section}{-1} \fi
   \@startsection {section}{1}{\z@}{-3.5ex plus -1ex minus 
   -.2ex}{2.3ex plus .2ex}{\Large\bf}}

\def\paragraph#1{{\bf #1\ }}

\newtheorem{lemma}{Lemma}[section]  

\newtheorem{theorem}[lemma]{Theorem}

\newtheorem{proposition}[lemma]{Proposition}

\title{Continuum limit of self-driven particles with orientation interaction} 
\author{P. Degond $^{(1)}$, S. Motsch$^{(1)}$} 
\date{} 
\begin{document}

\maketitle

\vspace{0.5 cm}

\begin{center}
(1)\, Institute of Mathematics of Toulouse 
UMR 5219 (CNRS-UPS-INSA-UT1-UT2),
Universit\'e Paul Sabatier,
118, route de Narbonne,  31062 Toulouse cedex, 
France \\
email: degond@mip.ups-tlse.fr, motsch@mip.ups-tlse.fr
\end{center}

\vspace{0.5 cm}
\begin{abstract}
We consider the discrete Couzin-Vicsek algorithm (CVA) \cite{Aldana_Huepe,couzin02:_collec_memor,Gregoire_Chate,Vicsek95:_novel}, which describes the interactions of individuals among animal societies such as fish schools. In this article, we propose a kinetic (mean-field) version of the CVA model and provide its formal macroscopic limit. The final macroscopic model involves a conservation equation for the density of the individuals and a non conservative equation for the director of the mean velocity and is proved to be hyperbolic. The derivation is based on the introduction of a non-conventional concept of a collisional invariant of a collision operator. 
\end{abstract}

\medskip
\noindent
{\bf Acknowledgements:} The first author wishes to thank E. Carlen and M. Carvalho for their interest in this work and their helpful suggestions. \\
$\mbox{}$ \\
Preprint of an article submitted for consideration in Mathematical Models and Methods in Applied Sciences (M3AS) © 2007 copyright World Scientific Publishing Company
http://www.worldscinet.com/m3as/m3as.shtml


\vskip 0.4cm

\setcounter{equation}{0}
\section{Introduction}
\label{intro}

The discrete Couzin-Vicsek algorithm (CVA) \cite{Aldana_Huepe,couzin02:_collec_memor,Gregoire_Chate,Vicsek95:_novel} has been proposed as a model for the interactions of individuals among animal societies such as fish schools. The individuals move with a velocity of constant magnitude. The CVA model describes in a discrete way the time evolution of the positions of the individuals and of their velocity angles measured from a reference direction. At each time step, the angle is updated to a new value given by the director of the average velocity of the neighbouring particles, with addition of noise. The positions are updated by adding the distance travelled during the time step by the fish in the direction speficied by its velocity angle. 

For the modeling of large fish schools which can reach up to several milion individuals, it may be more efficient to look for continuum like models, which describe the fish society by macroscopic variables (e.g.  mean density, mean velocity and so on). Several such phenomenological models exist (see e.g. \cite{mogilner99,topaz04:_swarm,Topaz_Bertozzi_Lewis}). Several attemps to derive continuum models from the CVA model are also reported in the literature \cite{KRZB,RBKZ,RKZB}, but the derivation and the mathematical 'qualities' of the resulting models have not been fully analyzed yet. One can also refer to  \cite{orsogna06:_self,mogilner03:_mutual} for related  models. An alternate model, the Persistent Turning Walker model, has been proposed in \cite{Gautrais_Theraulaz} on the basis of experimental measurements. Its large-scale dynamics is studied in \cite{DM_curvature}. Additional references on swarm aggregation and fish schooling can be found in \cite{camazine02:_self_organ_biolog_system}. Among other types of animal societies, ants have been the subject of numerous studies and the reader can refer (among other references) to \cite{jost07:_from, theraulaz02:_spatial}, and references therein.

In this work, we propose a derivation of a continuum model from a kinetic version of the CVA algorithm. For that purpose, we first rephrase the CVA model as a time continuous dynamical system (see section \ref{sec_CV}). Then, we pass to a mean-field version of this dynamical system (section \ref{sec_Mean_Field}). This mean field model consists of a kinetic equation of Fokker-Planck type with a force term resulting from the alignement interactions between the particles. More precisely, the mean-field model is written:
\begin{eqnarray} 
& & \varepsilon ( \partial_t f^\varepsilon + \omega \cdot \nabla_x f^\varepsilon ) = -  \nabla_\omega \cdot (F_0^\varepsilon f^\varepsilon) + d \Delta_\omega f^\varepsilon + O(\varepsilon^2),  
\label{FP_mf_eps_0} \\
& & F_0^\varepsilon (x, \omega, t) = \nu \, \, (\mbox{Id} - \omega \otimes \omega) 
\Omega^\varepsilon (x,t) , 
\label{Force_mf_eps_0} \\
& & \Omega^\varepsilon (x, t) = \frac{j^\varepsilon (x, t)}{|j^\varepsilon (x, t)|}, \quad \mbox{ and } \quad j^\varepsilon (x, t) = \int_{ \upsilon \in {\mathbb S}^2 } 
\upsilon \, f^\varepsilon (x, \upsilon,t) \, d\upsilon
 \, . 
\label{bar_omega_mf_eps_0} 
\end{eqnarray}
Here $f^\varepsilon(x,\omega, t)$ is the particle distribution function depending on the space variable $x \in {\mathbb R}^3$, the velocity direction $\omega \in {\mathbb S}^2$ and the time $t$. $d$ is a scaled diffusion constant and $F_0^\varepsilon (x, \omega, t)$ is the mean-field interaction force between the particles which depends on an interaction frequency $\nu$. This force tends to align the particles to the direction $\Omega^\varepsilon$ which is the director of the particle flux $j^\varepsilon$. the operators $\nabla_\omega \cdot$ and $\Delta_\omega$ are respectively the gradient and the Laplace-Beltrami operators on the sphere. The matrix $(\mbox{Id} - \omega \otimes \omega) $ is the projection matrix onto the normal plane to $\omega$. $\varepsilon \ll 1$ is a small parameter measuring the ratio of the microscopic length scale (the distance travelled between two interactions) to the size of the  observation domain. Here, the relevant scaling is a hydrodynamic scaling, which means that $\varepsilon$ also equals the ratio of the microscopic time scale (the mean time between interactions) to the macroscopic observation time.  

The 'hydrodynamic limit' $\varepsilon \to 0$ provides the large-scale dynamics of the CVA model (in its mean-field version (\ref{FP_mf_eps_0})-(\ref{bar_omega_mf_eps_0})). 
The goal of this paper is to (formally) investigate this limit. More precisely, the main result of this paper is the following theorem, which is proved in section \ref{sec_hydro}: 

\begin{theorem} (formal)
The limit $\varepsilon \to 0$ of $f^\varepsilon$ is given by $f^0 = \rho M_{\Omega}$ where $\rho = \rho(x,t) \geq 0$ is the total mass of $f^0$ and $\Omega = \Omega(x,t) \in {\mathbb S}^2$ is the director of its flux: 
\begin{eqnarray} 
& & \rho(x,t) = \int_{\omega \in {\mathbb S}^2} f^0(x,\omega, t) \, d\omega , \label{mass_0} \\
& & \Omega = \frac{j}{|j|} \, , \quad j(x,t) = 
\int_{\omega \in {\mathbb S}^2} f^0(x,\omega, t) \, \omega \, d\omega , \label{Omega_0}
\end{eqnarray}
and $M_{\Omega}$ is a given function of $\omega \cdot \Omega$ only depending on $\nu$ and $d$ which will be specified later on (see (\ref{M_def})). 
Furthermore, $\rho(x,t)$ and $\Omega(x,t)$ satisfy the following system of first order partial differential equations: 
\begin{eqnarray} 
& & \partial_t \rho + \nabla_x \cdot (c_1 \rho \Omega)  = 0.
\label{mass_eq_0} \\
& &  \rho \, \left( \partial_t \Omega + c_2 (\Omega \cdot \nabla) \Omega \right) +  \lambda  \,  (\mbox{Id} - \Omega \otimes \Omega) \nabla_x \rho = 0,
\label{Omega_eq_0} 
\end{eqnarray}
where the convection speeds $c_1$, $c_2$ and the interaction constant $\lambda$ will be specified in the course of the paper (see (\ref{c_1}) and (\ref{c_2_mu})). 
\label{theo_limit}
\end{theorem}

Hydrodynamic limits have first been developed in the framework of the Boltzmann theory of rarefied gases. The reader can refer to \cite{CIP,degond03:_macros_boltz,Sone} for recent viewpoints as well as to \cite{Caflisch,Dip_Lio,Yu} for major landmarks in its mathematical theory. Hydrodynamic limits have been recently investigated in traffic flow modeling \cite{aw02:_deriv,helbing01:_traff} as well as in supply chain research \cite{armbruster06,Deg_Rin}. 

From the viewpoint of hydrodynamic limits, the originality of theorem \ref{theo_limit} lies in the fact that the collision operator (i.e. the right-hand side of (\ref{FP_mf_eps_0})) has a three dimensional manifold of equilibria (parametrized by the density $\rho$ and the velocity director $\Omega$) but has only a one-dimensional set of collisional invariants (corresponding to mass conservation). Indeed, the interaction does not conserve momentum and one should not expect any collisional invariant related to that conservation. The problem is solved by introducing a broader class of collisional invariants, such that their integral (with respect to $\omega$) against the collision operator cancels only when the collision operator is applied to a subclass of functions. Here, a generalized class of collision invariants is associated with each direction $\Omega$ on the sphere and the corresponding subclass of functions have their flux in the direction of $\Omega$. We show that such generalized collision invariants exist and that they lead to (\ref{Omega_eq_0}). In section \ref{subsec_hyper}, we show that this system is hyperbolic. The detailed qualitative study of the system as well as numerical simulations will be the subject of future work. A summary of this work can be found in \cite{Degond_Motsch_CV_CRAS}. 

An important consequence of this result is that the large-scale dynamics of the CVA model does not present any phase transition, in contrast with the observations of \cite{Vicsek95:_novel}. Indeed, the equilibrium is unique (for given density and velocity director). Therefore, the model cannot exhibit any bi-stable behavior where shifts between two competing equilibria would trigger abrupt phase transitions, like in rod-like polymers (see e.g. \cite{LZZ} and references therein). Instead, the equilibrium gradually shifts from a collective one where all particles point in the same direction to an isotropic one as the diffusion constant $d$ increases from $0$ to infinity. Additionally, the hyperbolicity of the model does not allow lines of faults across unstable elliptic regions, like in the case of multi-phase mixtures or phase transitions in fluids or solids (see e.g. the review in \cite{Keyfitz} and references therein). 

With these considerations in mind, a phenomenon qualitatively resembling a phase transition could occur if the coefficients $c_1$, $c_2$ and $\lambda$ have sharp variations in some small range of values of the diffusion coefficient $d$. In this case, the model could undergo a rapid change of its qualitative features which would be reminiscent of a phase transition. One of our future goals is to verify or discard this possibility by numerically computing these constants. 

There are many questions which are left open. For instance, one question is about the possible influence of a limited range of vision in the backwards direction. In this case, the asymetry of the observation will bring more terms in the limit model. Similarly, one could argue that the angular diffusion should produce some spatial dissipation. Indeed, such dissipation phenomena are likely to occur if we retain the first order correction in the series expansion in terms of the small parameter $\varepsilon$ (the so-called Hilbert or Chapman-Enskog expansions, see e.g. \cite{degond03:_macros_boltz}). A deeper analysis is needed in order to determine the precise form of these diffusion terms. Another question concerns the possibility of retaining some of the non-local effects in the macroscopic model. It is likely that the absence of phase transition in the present model is related to the fact that the large-scale limit cancels most of the non-local effects (at least at leading order). The question whether retaining some non-locality effects in the macroscopic limit would allow the appearence of phase transitions at large scales would indeed reconcile the analytical result with the numerical observations.
A result in this direction obtained with methods from matrix recursions can be found in \cite{Cucker_Smale}. Finally, the alignement interaction is only one of the aspects of the Couzin model, which also involves repulsion at short scales and attraction at large scales. The incorporation of these effects would allow to build a complete continuum model which would account for all the important features of this kind of social interaction.

\setcounter{equation}{0}
\section{A time continuous version of the discrete Couzin-Vicsek algorithm}
\label{sec_CV}

The Couzin-Vicsek algorithm considers $N$ point particles in ${\mathbb R}^3$ labeled by $k \in \{1, \ldots N\}$ with positions $X_k^n$ at the discrete times $t^n = n \Delta t$. The  magnitude of the velocity is the same for all particles and is constant in time denoted by $c>0$.  The velocity vector is written $c \, \omega_k^n$ where $\omega_k^n$ belongs to the unit sphere ${\mathbb S}^2 = \{\omega $ \, s.t. $ |\omega|^2 = 1 \}$ of ${\mathbb R}^3$. 

The Couzin-Vicsek algorithm is a time-discrete algorithm that updates the velocities and positions of the particles at every time step $\Delta t$ according to the following rules.  

\medskip 
\noindent
(i) The particle position of the $k$-th particle at time $n$ is evolved according to: 
\begin{eqnarray}
& & X_k^{n+1} = X_k^n + c \, \omega_k^n \Delta t . \label{CV_pos} 
\end{eqnarray}

\medskip 
\noindent (ii) The velocity director of the $k$-th particle, $\omega_k^n$, is changed to the director $\bar \omega_k^n$ of the average velocity of the neighboring particles with addition of noise. This algorithm tries to mimic the behaviour of some animal species like fish, which tend to align with their neighbors. Noise accounts for the inaccuracies of the animal perception and cognitive systems. The neighborhood of the $k$-th particle is the ball centered at $X_k^n$ with radius $R>0$ and $\bar \omega_k^n$ is given by:
\begin{eqnarray}
& & \bar \omega_k^n = \frac{J_k^n}{|J_k^n|}, \quad J_k^n = 
\sum_{j,  \, |X_j^n - X_k^n| \leq R} \omega_j^n   . 
\label{CV_moy_3} 
\end{eqnarray}
In the Couzin-Vicsek algorithm, the space is 2-dimensional and the orientations are vectors belonging to the unit sphere ${\mathbb S}^1$ in ${\mathbb R}^2$.
One can write $\omega_k^n = e^{i \theta_k^n}$ with $\theta_k^n$ defined modulo $2 \pi$, and similarly  $\bar \omega_k^n = e^{i \bar \theta_k^n}$. In the original version of the algorithm, a uniform noise in a small interval of angles $[- \alpha, \alpha]$ is added, where $\alpha$ is a measure of the intensity of the noise. This leads to the following update for the phases: 
\begin{eqnarray}
& & \theta_k^{n+1} = \bar \theta_k^n + \hat \theta_k^n , 
\label{CV_phase_update} 
\end{eqnarray}
where $\hat \theta_k^n$ are independent identically distributed random variables with uniform distribution in $[- \alpha, \alpha]$. Then, $\omega_k^{n+1} = e^{i \theta_k^{n+1}}$. In \cite{Vicsek95:_novel}, Vicsek et al analyze the dynamics of this algorithm and experimentally demonstrate the existence of a threshold value $\alpha^*$. For $\alpha < \alpha^*$, a coherent dynamics appears after some time where all the particles are nearly aligned. On the other hand, if $\alpha > \alpha^*$, disorder prevails at all times.

Here, we consider a three dimensional version of the Couzin-Vicsek algorithm, of which the two-dimensional original version is a particular case. Of course, formula (\ref{CV_moy_3}) for the average remains the same in any dimension. For simplicity, we also consider a Gaussian noise rather than a uniformly distributed noise as in the original version of the algorithm. Therefore, our algorithm updates the velocity directors according to:  
\begin{eqnarray}
& & \omega_k^{n+1} = \hat \omega_k^n  , 
\label{CV_phase_update2} 
\end{eqnarray}
where $\hat \omega_k^n$ are random variables on the sphere centered at $\bar \omega_k^n$ with  Gaussian distributions of variance $\sqrt{2 D \Delta t}$ where $D$ is a supposed given coefficient. If the Gaussian noise is discarded, the evolution of the orientations is given by 
\begin{eqnarray}
& & \omega_k^{n+1} = \bar \omega_k^n  , 
\label{CV_phase_update3} 
\end{eqnarray}
where $\bar \omega_k^n$ is the average defined at (\ref{CV_moy_3}). 

Now, we would like to take the limit $\Delta t \to 0$ and find a time-continuous dynamics. To do so, we first consider the deterministic algorithm (\ref{CV_pos}), (\ref{CV_phase_update3}) and following \cite{RBKZ}, make some elementary remarks. First, because $|\omega_k^n| =  |\omega_k^{n+1}|$, we have $(\omega_k^{n+1} - \omega_k^n) (\omega_k^{n+1} + \omega_k^n) = 0$. Therefore, defining $\omega_k^{n+1/2} = (\omega_k^{n+1} + \omega_k^n)/2$ and using (\ref{CV_phase_update3}), we have the obvious relation: 
\begin{eqnarray} 
& & \frac{\omega_k^{n+1} - \omega_k^n}{\Delta t}  = \frac{1}{\Delta t} (\mbox{Id} - \omega_k^{n+1/2} \otimes \omega_k^{n+1/2}) 
(\bar \omega_k^n - \omega_k^n) , 
\label{CV_orthogonality} 
\end{eqnarray}
where Id denotes the Identity matrix and the symbol $\otimes$ denotes the tensor product of vectors. The matrix \, $\mbox{Id} - \omega_k^{n+1/2} \otimes \omega_k^{n+1/2}$ \, is the orthogonal projector onto the plane orthogonal to $\omega_k^{n+1/2}$. Relation (\ref{CV_orthogonality}) simply expresses that the vector $\omega_k^{n+1} - \omega_k^n$ belongs to that plane. 

Now, we let $\Delta t \to 0$. Then, the positions $X_k(t)$ and the orientations $\omega_k(t)$ become continuous functions of time. If we let $\Delta t \to 0$ in (\ref{CV_orthogonality}), the left-hand side obviously tends to $\partial \omega_k / \partial t$. The right hand side, however, does not seem to have an obvious limit. This is due to an improper choice of time scale. Indeed, if we run the clock twice as fast, the particles will interact twice as frequently. In the limit $\Delta t \to 0$, the number of interactions per unit of time is infinite and we should not expect to find anything interesting if we do not rescale the time. In order to have the proper time-scale for the model, we need to replace the tick of the clock $\Delta t$ by a typical interaction frequency $\nu$ of the particles under consideration. For instance, in the case of fish, $\nu^{-1}$ is the typical time-interval between two successive changes in the fish trajectory to accomodate the presence of other fish in the neighbourhood. 
Therefore, we start from a discrete algorithm defined by 
\begin{eqnarray} 
& & \frac{\omega_k^{n+1} - \omega_k^n}{\Delta t}  = \nu \, (\mbox{Id} - \omega_k^{n+1/2} \otimes \omega_k^{n+1/2}) 
(\bar \omega_k^n - \omega_k^n) , 
\label{CV_orthogonality_2} 
\end{eqnarray}
together with (\ref{CV_pos}) and in the limit $\Delta t \to 0$, we find the following continuous dynamical system: 
\begin{eqnarray}
& & \frac{d X_k}{dt} =  c \, \omega_k , \label{Cont_pos} \\
& & \frac{d \omega_k}{dt} = \nu \, (\mbox{Id} - \omega_k \otimes \omega_k) 
\bar \omega_k  , 
\label{Cont_orient} 
\end{eqnarray}
where we have used that $(\mbox{Id} - \omega_k \otimes \omega_k) \omega_k = 0$. 
If the Gaussian noise is retained, then, the limit $\Delta t \to 0$ of the discrete algorithm is the following Stochastic Differential Equation: 
\begin{eqnarray}
& & \frac{d X_k}{dt} =  c \, \omega_k , \label{Cont_pos_2} \\
& & d \omega_k =  (\mbox{Id} - \omega_k \otimes \omega_k) (\nu \, 
\bar \omega_k \, dt + \sqrt{2D} \, dB_t ), 
\label{Cont_orient_2} 
\end{eqnarray}
where $dB_t$ is a Brownian motion with intensity  $\sqrt{2D}$. Of course, this $\Delta t \to 0$ limit is formal but the convergence proof is outside the scope of the present paper. 

We slightly generalize this model by assuming that $\nu$ may depend on the angle between 
$\omega_k$ and $\bar \omega_k$, namely $\nu = \nu(\cos \theta_k)$, with $\cos \theta_k =  \omega_k \cdot \bar \omega_k$. Indeed, it is legitimate to think that the ability to turn is dependent on the target direction. If we are considering fish, the ability to turn a large angle is likely to be reduced compared to small angles. We will assume that $\nu(\cos \theta)$ is a smooth and bounded function of $\cos \theta$.

\setcounter{equation}{0}
\section{Mean-field model of the discrete Couzin-Vicsek algorithm}
\label{sec_Mean_Field}

We now consider the limit of a large number of particles $N \to \infty$. We first consider the case without Gaussian noise. For this derivation, we proceed e.g. like in \cite{Spohn}. We introduce the so-called empirical distribution $f^N(x,\omega,t)$ defined by: 
\begin{eqnarray} 
& & f^N(x,\omega,t) = \frac{1}{N} \sum_{k=1}^N \delta(x-X_k(t)) \, \delta(\omega, \omega_k(t))  . 
\label{Emp_meas} 
\end{eqnarray}
Here, the distribution $\omega \in {\mathbb S}^2 \to \delta(\omega, \omega')$ is defined by duality against a smooth function $\varphi$ by the relation:
$$ \langle \delta(\omega, \omega') , \varphi(\omega) \rangle = \varphi (\omega') . $$
We note that $\delta(\omega, \omega') \not = \delta (\omega-\omega')$ because the sphere ${\mathbb S}^2$ is not left invariant by the subtraction operation. However, we have similar properties of $\delta$ such as $\delta(\omega, \omega') = \delta(\omega', \omega)$ where this relation is interpreted as concerning a distribution on the product ${\mathbb S}^2 \times {\mathbb S}^2$. 

Then, it is an easy matter to see that $f^N$ satisfies the following kinetic equation 
\begin{eqnarray} 
& & \partial_t f^N + c \omega \cdot \nabla_x f^N + \nabla_\omega \cdot (F^N f^N) = 0 ,  
\label{Liouville} 
\end{eqnarray}
where $F^N (x, \omega, t)$ is an interaction force defined by:
\begin{eqnarray} 
& & F^N (x, \omega, t) = \nu(\cos \theta^N) \, (\mbox{Id} - \omega \otimes \omega)
\bar \omega^N , 
\label{Force_N} 
\end{eqnarray}
with $\cos \theta^N = \omega \cdot \bar \omega^N$ 
and $\bar \omega^N (x, \omega,  t)$ is the average orientation around $x$, given by: 
\begin{eqnarray} 
& & \bar \omega^N (x, \omega, t) = \frac{J^N(x,t)}{|J^N(x,t)|}, \quad J^N(x,t) = 
\sum_{j,  \, |X_j^n - x| \leq R} \omega_j^n \, . 
\label{bar_omega_N} 
\end{eqnarray}
If, by any chance, $J^N$ is equal to zero, we decide to assign to $\bar \omega^N (x,\omega, t)$ the value $\omega$ (which is the only way by which $\bar \omega^N (x,\omega, t)$ can depend on $\omega$). In the sequel, this convention will not be recalled but will be marked by showing the dependence of $\bar \omega$ upon $\omega$

We recall the expressions of the gradient and divergence operator on the sphere. Let $x=(x_1,x_2,x_3)$ be a cartesian coordinate system associated with an orthonormal basis $(e_1,e_2,e_3)$ and let $(\theta, \phi)$ be a spherical coordinate system associated with this basis, i.e. $x_1= \sin \theta \cos \phi$, $x_2=  \sin \theta \sin \phi$, $x_3= \cos \theta$. Let also $(e_\theta, e_\phi)$ be the local basis associated with the spherical coordinate system ; the vectors  $e_\theta$ and $e_\phi$ have the following coordinates in the cartesian basis: $e_\theta = (\cos \theta \cos \phi, \cos \theta \sin \phi, -\sin \theta)$, $e_\phi = (- \sin \phi, \cos \phi, 0)$. Let $f(\omega)$ be a scalar function and $A= A_\theta e_\theta + A_\phi e_\phi$ be a tangent vector field. Then: 
$$\nabla_\omega f =  \partial_\theta f \,  e_\theta + \frac{1}{\sin \theta} \, \partial_\phi f \, e_\phi, \quad \nabla_\omega \cdot A = \frac{1}{\sin \theta} \partial_\theta (A_\theta \sin \theta) + \frac{1}{\sin \theta} \partial_\phi A_\phi. $$
If the cartesian coordinate system is such that $e_3 =  \bar \omega^N$, then 
\begin{eqnarray} 
& & F^N = - \nu(\cos \theta) \, \sin \theta \, \, e_\theta. 
\label{Force_N_sph} 
\end{eqnarray}

Back to system (\ref{Liouville})-(\ref{bar_omega_N}), we note that relation (\ref{bar_omega_N}) can be written
\begin{eqnarray} 
& & \hspace{-1cm}
\bar \omega^N (x, \omega, t) = \frac{J^N(x,t)}{|J^N(x,t)|}, \quad J^N(x,t) = \int_{ |y - x| \leq R , \, \upsilon \in {\mathbb S}^2 } 
\upsilon f^N(y, \upsilon,t) \, dy \, d\upsilon  \, . 
\label{bar_omega_N_2} 
\end{eqnarray}
We will slightly generalize this formula and consider $\bar \omega^N (x, \omega, t)$ defined by the following relation: 
\begin{eqnarray} 
& & \hspace{-1cm}
\bar \omega^N (x, \omega, t) = \frac{J^N(x,t)}{|J^N(x,t)|}, \quad J^N(x,t) = 
\int_{ y \in {\mathbb R}^3 , \, \upsilon \in {\mathbb S}^2 } K(|x-y|) \, 
\upsilon \, f^N(y, \upsilon,t) \, dy \, d\upsilon \,  \, , 
\label{bar_omega_N_3} 
\end{eqnarray}
where $K(|x|)$ is the 'observation kernel' around each particle. Typically, in formula (\ref{bar_omega_N_2}), $K(|x|)$ is the indicator function of the ball centered at the origin and of radius $R$ but we can imagine more general kernels modeling the fact that the influence of the particles fades away with distance. We will assume that this function is smooth, bounded and tends to zero at infinity. 

Clearly, the formal mean-field limit of the particle system modeled by the kinetic system (\ref{Liouville}), (\ref{Force_N}), (\ref{bar_omega_N_3}) is given by the following system: 
\begin{eqnarray} 
& & \hspace{-1cm} \partial_t f + c \omega \cdot \nabla_x f + \nabla_\omega \cdot (F f) = 0 ,  
\label{Liouville_mf} \\
& & \hspace{-1cm} F (x, \omega, t) = \nu(\cos \bar \theta) \, (\mbox{Id} - \omega \otimes \omega) 
\bar \omega (x,\omega,t) , 
\label{Force_mf} \\
& & \hspace{-1cm} \bar \omega (x, \omega, t) = \frac{J(x,t)}{|J(x,t)|}, \quad J(x,t) =  \int_{ y \in {\mathbb R}^3 , \, \upsilon \in {\mathbb S}^2 } K(|x-y|) \, 
\upsilon \, f(y, \upsilon,t) \, dy \, d\upsilon \, \, , 
\label{bar_omega_mf} 
\end{eqnarray}
with $\cos \bar \theta = \omega \cdot \bar \omega$. 
It is an open problem to rigorously show that this convergence holds. For interacting particle system, a typical result is as follows (see e.g. \cite{Spohn}). Suppose that the empirical measure at time $t=0$ converges in the weak star topology of bounded measures towards a smooth function $f_I (x, \omega)$. Then,  $f^N (x, \omega,t)$ converges to the solution $f$ of (\ref{Liouville_mf})-(\ref{bar_omega_mf}) with initial datum $f_I$,  in the topology of continous functions of time on $[0,T]$ (for arbitrary $T>0$) with values in the space of bounded measures endowed with the weak star topology. We will admit that such a result is true (may be with some modifid functinal setting) and leave a rigorous convergence proof to future work. 

We will also admit that the mean-field limit of the stochastic particle system (\ref{Cont_pos_2}), (\ref{Cont_orient_2}) consists of the following Kolmogorov-Fokker-Planck equation 
\begin{eqnarray} 
& & \partial_t f + c \omega \cdot \nabla_x f + \nabla_\omega \cdot (F f) = D \Delta_\omega f ,  
\label{FP_mf} 
\end{eqnarray}
again coupled with (\ref{Force_mf}), (\ref{bar_omega_mf}) for the definition of $F$ and $\bar \omega$, and where $\Delta_\omega$ denotes the Laplace-Belltrami operator on the sphere: 
$$ \Delta_\omega f = \nabla_\omega \cdot \nabla_\omega f =   \frac{1}{\sin \theta} \partial_\theta ( \sin \theta \partial_\theta f )  + 
\frac{1}{\sin^2 \theta} \partial_{\phi \phi}  f  . $$

\setcounter{equation}{0}
\section{Hydrodynamic limit of the Mean-field Couzin-Vicsek model}
\label{sec_hydro}

\subsection{Scaling}
\label{subsec_scaling}

We are interested in the large time and space dynamics of the mean-field Fokker-Planck equation (\ref{FP_mf}), coupled with (\ref{Force_mf}), (\ref{bar_omega_mf}). 

So far, the various quantities appearing in the system have physical dimensions. We first introduce the characteristic physical units associated with the problem and scale the system to dimensionless variables. Let $\nu_0$ the typical interaction frequency scale. This means that $ \nu(\cos \theta) = \nu_0 \nu' (\cos \theta)$ with $\nu' (\cos \theta) = O(1)$ in most of the range of $\cos \theta$. We now introduce related time and space scales $t_0$ and $x_0$ as follows: $t_0 = \nu_0^{-1}$ and $x_0 = c t_0 = c / \nu_0$. 
This choice means that the time unit is the mean time between interactions and the space unit is the mean distance traveled by the particles between interactions. We introduce the dimensionless diffusion coefficient $d = D/\nu_0$. Note that $D$ has also the dimension of a frequency so that $d$ is actually dimensionless. We also introduce a scaled observation kernel $K'$ such that $K(x_0 |x'|) = K'(|x'|)$. Typically, if $K$ is the indicator function of the ball of radius $R$, $K'$ is the indicator of the ball of radius $R' = R/x_0$. The assumption that the interaction is non local means that $R'=O(1)$. In other words, the observation radius is of the same order as the mean distance travelled by the particles between two interactions. This appears consistent with the behaviour of a fish, but would probably require more justifications. In the present work, we shall take this fact for granted. 

Let now $t' = t/t_0$, $x' = x/x_0$ the associated dimensionless time and space variables.
Then, system (\ref{FP_mf}), coupled with (\ref{Force_mf}), (\ref{bar_omega_mf}) is written in this new system of units (after dropping the primes for the sake of clarity): 
\begin{eqnarray} 
& & \hspace{-1cm} \partial_t f + \omega \cdot \nabla_x f + \nabla_\omega \cdot (F f) = d \Delta_\omega f ,  
\label{FP_mf_ad} \\
& & \hspace{-1cm} F (x, \omega, t) = \nu(\cos \bar \theta) \, (\mbox{Id} - \omega \otimes \omega) 
\bar \omega (x,\omega,t) , \quad \mbox{with} \quad  \cos \bar \theta = \omega \cdot \bar \omega , 
\label{Force_mf_ad} \\
& & \hspace{-1cm} \bar \omega (x, \omega, t) = \frac{J(x,t)}{|J(x,t)|}, \quad J(x,t) =  \int_{ y \in {\mathbb R}^3 , \, \upsilon \in {\mathbb S}^2 } K(|x-y|) \, 
\upsilon \, f(y, \upsilon,t) \, dy \, d\upsilon  \, , 
\label{bar_omega_mf_ad} 
\end{eqnarray}
The system now depends on only one dimensionless parameter $d$ and two dimensionless functions which describe the behaviour of the fish: $\nu(\cos \bar \theta)$ and $K(x)$, which are all supposed to be of order $1$.

Up to now, the system has been written at the microscopic level, i.e. at time and length scales which are characteristic of the dynamics of the individual particles. Our goal is now to investigate the dynamics of the system at large time and length scales compared with the scales of the individuals. For this purpose, we adopt new time and space units $\tilde t_0 = t_0 / \varepsilon$, $\tilde x_0 = x_0 / \varepsilon$ with $\varepsilon \ll 1$. Then, a set of new dimensionless variables is introduced $\tilde x = \varepsilon x$, $\tilde t = \varepsilon t$. In this new set of variables, the system is written (again, dropping the tildes for clarity): 
\begin{eqnarray} 
& & \hspace{-1cm} \varepsilon ( \partial_t f^\varepsilon + \omega \cdot \nabla_x f^\varepsilon ) = -  \nabla_\omega \cdot (F^\varepsilon f^\varepsilon) + d \Delta_\omega f^\varepsilon ,  
\label{FP_mf_large} \\
& & \hspace{-1cm} F^\varepsilon (x, \omega, t) = \nu(\omega \cdot \bar \omega^\varepsilon) \, \, (\mbox{Id} - \omega \otimes \omega) 
\bar \omega^\varepsilon (x,\omega,t) , 
\label{Force_mf_large} \\
& & \hspace{-1cm} \bar \omega^\varepsilon (x, \omega, t) = \frac{J^\varepsilon(x,t)}{|J^\varepsilon(x,t)|}, \quad J^\varepsilon(x,t) = \int_{ y \in {\mathbb R}^3 , \, \upsilon \in {\mathbb S}^2 } K \left( \left| \frac{x-y}{\varepsilon} \right| \right) \, 
\upsilon \, f^\varepsilon(y, \upsilon,t) \, dy \, d\upsilon  \, , 
\label{bar_omega_mf_large} 
\end{eqnarray}
Our goal in this paper is to investigate the formal limit $\varepsilon \to 0$ of this problem.

Our first task, performed in the following lemma,  is to provide an expansion of $\bar \omega^\varepsilon$ in terms of $\varepsilon$. 

\begin{lemma} We have the expansion: 
\begin{eqnarray} 
& & \bar \omega^\varepsilon (x, \omega, t) = \Omega^\varepsilon (x, t) + O(\varepsilon^2)
 \, ,
\label{bar_omega_expan} 
\end{eqnarray}
where 
\begin{eqnarray} 
& & \Omega^\varepsilon (x, t) = \frac{j^\varepsilon (x, t)}{|j^\varepsilon (x, t)|}, \quad \mbox{ and } \quad j^\varepsilon (x, t) = \int_{ \upsilon \in {\mathbb S}^2 } 
\upsilon \, f^\varepsilon (x, \upsilon,t) \, d\upsilon
 \, .
\label{Omega} 
\end{eqnarray}
\label{lem_bar_omega_expan}
\end{lemma}

The proof of this lemma is elementary, and is omitted. That the remainder in (\ref{bar_omega_expan}) is of order $\varepsilon^2$ is linked with the fact that the observation kernel is isotropic. If an anisotropic kernel had been chosen, such as one favouring observations in the forward direction, then a term of order $\varepsilon$ would have been obtained. This additional term would substantially change the dynamics. We leave this point to future work. 

The quantity $j^\varepsilon (x, t)$ is the particle flux. We will also use the density, which is defined as a moment of $f$ as well:
\begin{eqnarray} 
& & \rho^\varepsilon (x, t) = \int_{ \upsilon \in {\mathbb S}^2 } 
 f^\varepsilon (x, \upsilon,t) \, d\upsilon
 \, .
\label{density} 
\end{eqnarray}

Thanks to lemma \ref{lem_bar_omega_expan}, system (\ref{FP_mf_large})-(\ref{bar_omega_mf_large}) is written
\begin{eqnarray} 
& & \varepsilon ( \partial_t f^\varepsilon + \omega \cdot \nabla_x f^\varepsilon ) = -  \nabla_\omega \cdot (F_0^\varepsilon f^\varepsilon) + d \Delta_\omega f^\varepsilon + O(\varepsilon^2),  
\label{FP_mf_eps} \\
& & F_0^\varepsilon (x, \omega, t) = \nu(\omega \cdot \Omega^\varepsilon) \, \, (\mbox{Id} - \omega \otimes \omega) 
\Omega^\varepsilon (x,t) , 
\label{Force_mf_eps} \\
& & \Omega^\varepsilon (x, t) = \frac{j^\varepsilon (x, t)}{|j^\varepsilon (x, t)|}, \quad \mbox{ and } \quad j^\varepsilon (x, t) = \int_{ \upsilon \in {\mathbb S}^2 } 
\upsilon \, f^\varepsilon (x, \upsilon,t) \, d\upsilon
 \, . 
\label{bar_omega_mf_eps} 
\end{eqnarray}
We note that observing the system at large scales makes the interaction local and that this interaction tends to align the particle velocity to the direction of the local particle flux. This interaction term is balanced at leading order by the diffusion term which tends to spread the particles isotropically on the sphere. Obviously, an equilibrium distribution results from the balance of these two antogonist phenomena. 

In the remainder of the paper, we write $F[f^\varepsilon]$ for $F_0^\varepsilon$. We introduce the operator 
\begin{eqnarray} 
& & Q(f) = -  \nabla_\omega \cdot (F[f] f) + d \Delta_\omega f ,  
\label{Q_def} \\
& & F[f] = \nu \, \, (\mbox{Id} - \omega \otimes \omega) \Omega[f] , 
\label{Force_def} \\
& & \Omega[f] = \frac{j[f]}{|\, \,  j[f]\, \,  |}, \quad \mbox{ and } \quad j[f] = \int_{ \omega \in {\mathbb S}^2 } 
\omega \, f \, d\omega
 \, . 
\label{Omega_def} 
\end{eqnarray}
We note that $\Omega[f]$ is a non linear operator of $f$, and so are  $F[f]$ and $Q(f)$. 
In the remainder, we will always suppose that $f$ is as smooth and integrable as necessary. We leave the question of finding the appropriate functional framework to forthcoming work. 

The operator $Q$ acts on the angle variable $\omega$ only and leaves the other variables $x$ and $t$ as parameters. Therefore, it is legitimate to study the properties of $Q$ as an operator acting on functions of $\omega$ only. This is the task performed in the following section.

\subsection{Properties of $Q$}
\label{subsec_prop_Q}

We begin by looking for the equilibrium solutions, i.e. the functions $f$ which cancel $Q$. Let $\mu = \cos \theta$. We denote by $\sigma(\mu)$ an antiderivative of $\nu(\mu)$, i.e. $(d \sigma / d \mu)(\mu) = \nu(\mu)$. We define 
\begin{eqnarray} 
& &  M_{\Omega}(\omega) = C \exp(\frac{\sigma(\omega \cdot \Omega)}{d}), \quad \int M_{\Omega}(\omega) \, d \omega =1 
 \, . 
\label{M_def} 
\end{eqnarray}
The constant $C$ is set by the normalization condition (second equality of (\ref{M_def}))~; it depends only on $d$ and on the function $\sigma$ but not on $\Omega$.

We have the following:
\begin{lemma}
(i) The operator $Q$ can be written as 
\begin{eqnarray} 
& &  Q(f) = d \, \, \nabla_\omega \cdot \left[ M_{\Omega[f]} \nabla_\omega \left( \frac{f}{M_{\Omega[f]}} \right) \right] ,    
\label{Q_self}
\end{eqnarray}
and we have 
\begin{eqnarray} 
& &  \hspace{-1cm} H(f) := \int_{\omega \in {\mathbb S}^2}  Q(f) \frac{f}{M_{\Omega[f]}} \, d\omega = - d \, \, \int_{\omega \in {\mathbb S}^2} M_{\Omega[f]} \left| \nabla_\omega \left( \frac{f}{M_{\Omega[f]}} \right) \right|^2 \, d\omega \leq 0 .     
\label{Q_entrop}
\end{eqnarray}

\noindent
(ii) The equilibria, i.e. the functions $f(\omega)$ such that $Q(f) = 0$ form a three-dimensional manifold ${\mathcal E}$ given by 
\begin{eqnarray} 
& & {\mathcal E} = \{ \rho M_{\Omega}(\omega)\quad | \quad \rho \in {\mathbb R}_+, \quad \Omega \in {\mathbb S}^2
\} 
 \, , 
\label{equi_mani} 
\end{eqnarray}
and $\rho$ is the total mass while $\Omega$ is the director of the flux of $\rho M_{\Omega}(\omega)$, i.e.
\begin{eqnarray} 
& &  \int_{\omega \in {\mathbb S}^2} \rho \, M_{\Omega}(\omega) \, d\omega = \rho \label{Max_mass} \\
& & \Omega = \frac{j[\rho M_{\Omega}]}{|\, j[\rho M_{\Omega}]\, |}  \, , \quad j[\rho M_{\Omega}] = \int_{\omega \in {\mathbb S}^2} \rho M_{\Omega}(\omega) \, \omega \,  d\omega   .   
\label{Max_flux}
\end{eqnarray}
Furthermore, $H(f) =0$ if and only if $f = \rho M_{\Omega}$ for arbitrary $\rho \in {\mathbb R}_+$ and $\Omega \in {\mathbb S}^2$. 
\label{lem_equi}
\end{lemma}

The function $\sigma$ being an increasing function of $\mu$ (since $\nu >0$), $M_\Omega$ is maximal for $\omega \cdot \Omega = 1$, i.e.  for $\omega$ pointing in the direction of $\Omega$. Therefore, $\Omega$ plays the same role as the average velocity of the classical Maxwellian of gas dynamics. The role of the temperature is played by the normalized diffusion constant $d$~: it measures the 'spreading' of the equilibrium about the average direction $\Omega$. Here the temperature is fixed by the value of the diffusion constant, in contrast with classical gas dynamics where the temperature is a thermodynamical variable whose evolution is determined by the energy balance equation. 

An elementary computation shows that the flux can be written
\begin{eqnarray} 
& &  j[\rho M_{\Omega}] = \langle \cos \theta \rangle_M \, \rho \Omega , \label{flux}
\end{eqnarray}
where for any function $g(\cos \theta)$, the symbol $\langle g(\cos \theta) \rangle_M$ denotes the average of $g$ over the probability distribution $M_\Omega$, i.e.
\begin{eqnarray} 
& &  \langle g(\cos \theta) \rangle_M = \int M_\Omega(\omega) g(\omega \cdot \Omega) \, d\omega = \frac{\int_0^\pi g(\cos \theta) \exp ( \frac{\sigma(\cos \theta)}{d})  \, 
\sin \theta \, d \theta}{\int_0^\pi \exp ( \frac{\sigma(\cos \theta)}{d}) \,   \sin \theta \, d \theta}
. \label{brackets}
\end{eqnarray}
We note that $\langle g(\cos \theta) \rangle_M$ does not depend on $\Omega$ but depends on $d$. In particular, $\langle g(\cos \theta) \rangle_M \to g(1)$ when $d \to 0$ while $\langle g(\cos \theta) \rangle_M \to \bar g$, the arithmetic average of $g$ over the sphere, when $d \to \infty$ (with $\bar g = \int g(\omega \cdot \Omega) \, d\omega = \frac{1}{2} \int_0^\pi g(\cos \theta)  \, 
\sin \theta \, d \theta$). Therefore, $\langle \cos \theta \rangle_M \to 1$ when $d \to 0$ and $\langle \cos \theta \rangle_M \to 0$ when $d \to \infty$. For a large diffusion, the equilibrium is almost isotropic and the magnitude of the velocity tends to zero while for a small diffusion, the distribution is strongly peaked in the forward direction and the magnitude of the velocity tends to $1$, which is the velocity of the individual particles.

\medskip
\noindent
{\bf Proof of lemma \ref{lem_equi}:} To prove (i), we introduce  a reference frame such that $e_3 = \Omega[f]$. In spherical coordinates, we have 
\begin{eqnarray} 
& &  M_{\Omega[f]}(\omega(\theta, \phi)) = C \exp( d^{-1} \sigma(\cos \theta)).    
\label{Q_zero_4} 
\end{eqnarray}
Therefore, 
\begin{eqnarray} 
 \nabla_\omega (\ln M_{\Omega[f]} ) &=& \nabla_\omega [ \, \ln \{ \, C \exp( d^{-1} \sigma(\cos \theta)) \, \} \, ] \nonumber \\
&=& d^{-1} \nabla_\omega (\sigma(\cos \theta)) \nonumber \\
&=& - d^{-1} \nu(\cos \theta) \sin \theta \, e_\theta \nonumber \\
&=& d^{-1} F[f] 
\, ,    
\label{Q_zero_5} 
\end{eqnarray}
where ln denotes the logarithm and the last equality results from (\ref{Force_N_sph}). 
Then, we deduce that  
\begin{eqnarray} 
d \, \, \nabla_\omega \cdot \left[ M_{\Omega[f]} \nabla_\omega \left( \frac{f}{M_{\Omega[f]}} \right) \right] &=&   
d \, \, \nabla_\omega \cdot \left[ \nabla_\omega f - f \nabla_\omega (\ln M_{\Omega[f]} ) \right] \nonumber \\
& = & d \Delta_\omega f - \nabla_\omega \cdot (F[f] f) = Q(f) . 
\label{Q_self_2}
\end{eqnarray}
(\ref{Q_entrop}) follows directly from (\ref{Q_self}) and Stokes theorem. 

(ii) follows directly from (i). If $Q(f)=0$, then $H(f)=0$. But $H(f)$ is the integral of a non-negative quantity and can be zero only if this quantity is identically zero, which means $f = \rho M_{\Omega[f]}$ for a conveniently chosen $\rho$. Since $\Omega[f]$ can be arbitrary, the result follows. The remaining statements are obvious. 
\endproof

Our task now is to determine the collision invariants of $Q$, i.e. the functions $\psi(\omega)$ such that 
\begin{eqnarray} 
\int_{\omega \in {\mathbb S}^2} Q(f) \, \psi \, d \omega = 0 , \quad \forall f . 
\label{Q_coll_invar}
\end{eqnarray}
Using (\ref{Q_self}), this equation can be rewritten as 
\begin{eqnarray} 
\int_{\omega \in {\mathbb S}^2} \frac{f}{M_{\Omega[f]}} \nabla_\omega \cdot ( M_{\Omega[f]} \nabla_\omega \psi ) \, d \omega = 0 , \quad \forall f . 
\label{Q_coll_invar_2}
\end{eqnarray}
Clearly, if $\psi =$ Constant, $\psi$ is a collisional invariant. On the other hand, there is no other obvious conservation relation, since momentum is  not conserved by the interaction operator. The constants span a one-dimensional function space, while the set of equilibria is a three-dimensional manifold. So, we need to find some substitute to the notion of collisional invariant, otherwise, in the limit $\varepsilon \to 0$, the problem will be under-determined, and in particular, we will lack an equation for  $\Omega$ (appearing in the expression of the equilibrium). 

To solve the problem, we slightly change the viewpoint. We fix $\Omega \in {\mathbb S}^2$ arbitrarily, and we ask the problem of finding all $\psi$'s which are collisional invariants of $Q(f)$ for all $f$ with director $\Omega[f] = \Omega$. Such a function $\psi$ is not a collisional invariant in the strict sense, because (\ref{Q_coll_invar}) is valid for all $f$ but only for a subclass of $f$. But this weaker concept of a collisional invariant is going to suffice for our purpose. 
So, for fixed $\Omega$, we want to find all $\psi$'s such that 
\begin{eqnarray} 
\int_{\omega \in {\mathbb S}^2} \frac{f}{M_{\Omega}} \nabla_\omega \cdot ( M_{\Omega} \nabla_\omega \psi ) \, d \omega = 0 , \quad \forall f \, \mbox{ such that } \, \Omega[f] = \Omega . 
\label{Q_coll_invar_3}
\end{eqnarray}
Now, saying that $\Omega[f] = \Omega$ is equivalent to saying that $j[f]$ is aligned with $\Omega[f]$, or again to 
\begin{eqnarray} 
0 = \Omega \times j[f]  = \int_{\omega \in {\mathbb S}^2} f \, (\Omega \times \omega) \, d \omega . 
\label{Q_coll_constraint}
\end{eqnarray}
This last formula can be viewed as a linear constraint and, introducing the Lagrange multiplier $\beta$ of this constraint, $\beta$ being a vector normal to $\Omega$, we can restate the problem of finding the 'generalized' collisional invariants (\ref{Q_coll_invar_3}) as follows: Given $\Omega \in {\mathbb S}^2$, find all $\psi$'s such that there exist $\beta \in {\mathbb R}^3$ with $\Omega \cdot \beta = 0$, and 
\begin{eqnarray} 
\int_{\omega \in {\mathbb S}^2} \frac{f}{M_{\Omega}} \left\{ \nabla_\omega \cdot ( M_{\Omega} \nabla_\omega \psi ) - \beta \cdot (\Omega \times \omega) M_{\Omega} \right\} \, d \omega = 0 , \quad \forall f  . 
\label{Q_coll_invar_4}
\end{eqnarray}
Now, (\ref{Q_coll_invar_4}) holds for all $f$ without constraint and immediately leads to the following problem for $\psi$: 
\begin{eqnarray} 
\nabla_\omega \cdot ( M_{\Omega} \nabla_\omega \psi ) = \beta \cdot (\Omega \times \omega) M_{\Omega} . 
\label{Q_coll_invar_5}
\end{eqnarray}

The problem defining $\psi$ is obviously linear, so that the set ${\mathcal C}_\Omega$ of generalized collisional invariants associated with the vector $\Omega$ is a vector space.
It is convenient to introduce a cartesian basis $(e_1, e_2, \Omega)$ and the associated spherical coordinates $(\theta, \phi)$. Then $\beta = (\beta_1, \beta_2, 0)$	and $\beta \cdot (\Omega \times \omega) = (- \beta_1 \sin \phi + \beta_2 \cos \phi) \sin \theta$.  Therefore, we can successively solve  for $\psi_1$ and $\psi_2$, the solutions of (\ref{Q_coll_invar_5}) with right-hand sides respectively equal to $- \sin \phi \sin \theta M_\Omega$ and $\cos \phi \sin \theta M_\Omega$. 

We are naturally looking for solutions in an $L^2({\mathbb S}^2)$ framework, since $\psi$ is aimed at constructing marcroscopic quantities by integration against $f$ with respect to $\omega$. Therefore, one possible framework is to look for both $f$ and $\psi$ in $L^2({\mathbb S}^2)$ to give a meaning to these macroscopic quantities. We state the following lemma: 

\begin{lemma}
Let $\chi \in L^2({\mathbb S}^2)$ such that $\int \chi \, d\omega = 0$. The problem 
\begin{eqnarray} 
\nabla_\omega \cdot ( M_{\Omega} \nabla_\omega \psi ) = \chi , 
\label{CI_elliptic}
\end{eqnarray}
has a unique weak solution in the space ${\stackrel{\circ}{H^1}}({\mathbb S}^2)$, the quotient of the space $H^1({\mathbb S}^2)$ by the space spanned by the constant functions, endowed with the quotient norm. 
\label{lem_exist}
\end{lemma}

\medskip
\noindent
{\bf Proof:} We apply the Lax-Milgram theorem to the following variational formulation of (\ref{CI_elliptic}): 
\begin{eqnarray} 
\int_{\omega \in {\mathbb S}^2}  M_{\Omega} \nabla_\omega \psi \cdot \nabla_\omega \varphi \, d\omega = \int_{\omega \in {\mathbb S}^2} \chi \varphi \, d \omega, 
\label{CI_var_form}
\end{eqnarray}
for all $\varphi \in {\stackrel{\circ}{H^1}}({\mathbb S}^2)$. The function $M_\Omega$ is bounded from above and below on ${\mathbb S}^2$, so the bilinear form at the left-hand side is continous on ${\stackrel{\circ}{H^1}}({\mathbb S}^2)$. The fact that the average of $\chi$ over ${\mathbb S}^2$ is zero ensures that the right-hand side is a continuous linear form on ${\stackrel{\circ}{H^1}}({\mathbb S}^2)$. The coercivity of the bilinear form is a consequence of the Poincare inequality: $\exists C>0$ such that $\forall \psi \in {\stackrel{\circ}{H^1}}({\mathbb S}^2)$: 
\begin{eqnarray} 
|\psi|_{H^1} \,  \geq \, C ||\psi||_{{\stackrel{\circ}{L^2}}} \, := \, C \min_{K \in {\mathbb R}} ||\psi + K||_{L^2} \, ,  
\label{CI_Poincare}
\end{eqnarray}
where $|\psi|_{H^1}$ is the $H^1$ semi-norm. We note that the Poincare inequality would not hold without taking the quotient. \endproof

So, to each of the right-hand sides $\chi = - \sin \phi \sin \theta M_\Omega$ or $\chi = \cos \phi \sin \theta M_\Omega$ which have zero average on the sphere, there exist solutions  $\psi_1$ and $\psi_2$ respectively (unique up to constants) of problem (\ref{CI_elliptic}). We single out unique solutions by requesting that $\psi_1$ and $\psi_2$ have zero average on the sphere: $\int \psi_k \, d\omega = 0$, $k=1,2$. We can state the following corollary to lemma \ref{lem_exist}: 

\begin{proposition}
The set  ${\mathcal C}_\Omega$ of generalized collisional invariants associated with the vector $\Omega$ which belong to $H^1({\mathbb S}^2)$ is a three dimensional vector space ${\mathcal C}_\Omega = \mbox{Span} \{ 1, \psi_1, \psi_2 \}$ 
\label{lem_coll_invar}
\end{proposition}

More explicit forms for $\psi_1$ and $\psi_2$ can be found. By expanding in Fourier series with respect to $\phi$, we easily see that 
\begin{eqnarray} 
& & \psi_1 = - g(\cos \theta) \sin \phi,  \quad \psi_2 = g(\cos \theta) \cos \phi , \label{psi_1_2} 
\end{eqnarray}
where $g(\mu)$ is the unique solution of the elliptic problem on $[-1,1]$:
\begin{eqnarray} 
& & - (1-{\mu}^2) \partial_\mu ( e^{\sigma(\mu)/d} (1-{\mu}^2) \partial_\mu g ) + e^{\sigma(\mu)/d} g = - (1-{\mu}^2)^{3/2} e^{\sigma(\mu)/d}
. \label{g} 
\end{eqnarray}
We note that no boundary condition is needed to specify $g$ uniquely since the operator at the left-hand side of (\ref{g}) is degenerate at the boundaries $\mu = \pm 1$. Indeed, it is an easy matter, using again Lax-Milgram theorem, to prove that problem (\ref{g}) has a unique solution in the weighted $H^1$ space $V$ defined by 
$$ V = \{ g \, | \, (1-\mu^2)^{-1/2} g \in L^2(-1,1), \quad (1-\mu^2)^{1/2} \partial_\mu g \in L^2(-1,1) \} . $$
Furthermore, the Maximum Principle shows that $g$ is non-positive. 

For convenience, we introduce  $h(\mu) = (1-\mu^2)^{-1/2} g \in L^2(-1,1) $ or equivalently $h(\cos \theta) = g(\cos \theta) / \sin \theta$.
We then define 
\begin{eqnarray} 
& & \vec \psi(\omega) = (\Omega \times \omega) \, h(\Omega \cdot \omega) = \psi_1 e_1 + \psi_2 e_2 \, 
. \label{psi} 
\end{eqnarray}
$\vec \psi$ is the vector generalized collisional invariant associated with the direction $\Omega$.

\subsection{Limit $\varepsilon \to 0$}
\label{subsec_limit}

The goal of this section is to prove theorem \ref{theo_limit}. 

Again, we suppose that all functions are as regular as needed and that all convergences are as strong as needed. The rigorous proof of this convergence result is outside the scope of this article. 

We start with eq. (\ref{FP_mf_eps}) which can be written 
\begin{eqnarray} 
& & \varepsilon ( \partial_t f^\varepsilon + \omega \cdot \nabla_x f^\varepsilon ) = Q( f^\varepsilon) + O(\varepsilon^2).   
\label{FP_mf_2} 
\end{eqnarray}
We suppose that $f^\varepsilon \to f$ when $\varepsilon \to 0$. Then, from the previous equation, $Q(f^\varepsilon) = O(\varepsilon)$ and we deduce that $Q(f) = 0$. By lemma 
\ref{lem_equi}, $f = \rho M_\Omega$, with $\rho \geq 0$ and $\Omega \in {\mathbb S}^2$. Now, since $Q$ operates on the variable $\omega$ only, this limit does not specify the dependence of $f$ on $(x,t)$, and consequently, $\rho$ and $\Omega$ are functions of $(x,t)$. 

To find this dependence, we use the generalized collisional invariants. First, we consider the constant collisional invariants, which merely means that we integrate (\ref{FP_mf_2}) with respect to $\omega$. We find the continuity equation 
\begin{eqnarray} 
& & \partial_t \rho^\varepsilon + \nabla_x \cdot j^\varepsilon  = 0,  
\label{mass_eq_eps} 
\end{eqnarray}
where $\rho^\varepsilon$ and $j^\varepsilon$ are the density and flux as defined above. It is an easy matter to realize that the right-hand side is exactly zero (and not $O(\varepsilon^2)$). In the limit $\varepsilon \to 0$,  $\rho^\varepsilon \to \rho$ and $j^\varepsilon \to j = c_1 \rho \Omega$ with 
\begin{eqnarray} 
& & c_1 = \langle \cos \theta \rangle_M,
\label{c_1} 
\end{eqnarray}
and we get  
\begin{eqnarray} 
& & \partial_t \rho + \nabla_x \cdot (c_1 \rho \Omega)  = 0.
\label{mass_eq} 
\end{eqnarray}

\medskip
Now, we multiply (\ref{FP_mf_2}) by $\vec \psi^\varepsilon = h(\omega \cdot \Omega[f^\varepsilon]) \, (\Omega[f^\varepsilon] \times \omega)$,  integrate with respect to $\omega$ and take the limit $\varepsilon \to 0$. We note that $\Omega[f^\varepsilon] \to \Omega$ and that $\vec \psi^\epsilon$ is smooth enough (given the functional spaces used for the existence theory), and consequently, 
$\vec \psi^\varepsilon \to \vec \psi = h(\omega \cdot \Omega) \, (\Omega \times \omega)$. Therefore, in the limit $\varepsilon \to 0$, we get:
\begin{eqnarray} 
& & \hspace{-1cm} \Omega \times X = 0 \, , \quad X:= \int_{\omega \in {\mathbb S}^2} ( \partial_t (\rho M_\Omega) + \omega \cdot \nabla_x (\rho M_\Omega)) \, h(\omega \cdot \Omega) \, \omega \, d \omega.   
\label{vec_psi_eq} 
\end{eqnarray}
Saying that $\Omega \times X = 0$ is equivalent to saying that the projection of $X$ onto the plane normal to $\Omega$ vanishes or in other words, that 
\begin{eqnarray} 
& & \hspace{-1cm} (\mbox{Id} - \Omega \otimes \Omega) X = 0 \, . 
\label{(Id-Omega)_X} 
\end{eqnarray}
This is the equation that we need to make explicit in order to find the evolution equation for $\Omega$.

Elementary differential geometry gives the derivative of $M_{\Omega}$ with respect to $\Omega$ acting on a tangent vector $d \Omega$ to the sphere as follows: 
\begin{eqnarray} 
& & \frac{\partial M_{\Omega}}{\partial \Omega} (d\Omega) = d^{-1} \nu(\omega \cdot \Omega) \, (\omega \cdot d\Omega) \, M_{\Omega}.
\label{deriv_M} 
\end{eqnarray}
We deduce that 
\begin{eqnarray} 
& & \partial_t (\rho M_{\Omega})  = M_{\Omega} \,  ( \partial_t \rho + d^{-1} \nu \, \rho  \, (\omega \cdot \partial_t \Omega)), 
\label{dt_M} \\
& & (\omega \cdot \nabla_x) (\rho M_{\Omega})  = M_{\Omega} \, ( (\omega \cdot \nabla_x) \rho + d^{-1} \nu \, \rho  \, \omega \cdot ((\omega \cdot \nabla_x) \Omega)). 
\label{omdx_M} 
\end{eqnarray}
Combining these two identities, we get: 
\begin{eqnarray} 
& &  \partial_t (\rho M_\Omega) + \omega \cdot \nabla_x (\rho M_\Omega) = \nonumber \\
& & \hspace{2cm}  = M_\Omega \left[ \partial_t \rho + \omega \cdot  \nabla_x \rho + d^{-1} \nu \rho ( \,  \omega \cdot \partial_t \Omega + (\omega \otimes \omega) : \nabla_x \Omega \,  ) \, \right] , 
\label{dt+omdx_M} 
\end{eqnarray}
where the symbol '$:$' denotes the contracted product of two tensors (if $A = (A_{ij})_{i,j=1, \ldots, 3}$ and $B = (B_{ij})_{i,j=1, \ldots, 3}$ are two tensors, then $A:B = \sum_{i,j=1, \ldots, 3} A_{ij} B_{ij}$) and $\nabla_x \Omega$ is the gradient tensor of the vector $\Omega$: $(\nabla_x \Omega)_{ij} = \partial_{x_i} \Omega_j$ . Therefore, the vector $X$, is given by: 
\begin{eqnarray} 
X= \int_{\omega \in {\mathbb S}^2}  \left[ \partial_t \rho + \omega \cdot  \nabla_x \rho + d^{-1} \nu \rho ( \,  \omega \cdot \partial_t \Omega + (\omega \otimes \omega) : \nabla_x \Omega \,  ) \, \right] \, \omega \, h \, M_\Omega \, d \omega
\label{X} 
\end{eqnarray}
The four terms in this formula, denoted by $X_1$ to $X_4$, are computed successively using spherical coordinates $(\theta, \phi)$ associated with a cartesian basis $(e_1, e_2, \Omega)$ where $e_1$ and $e_2$ are two vectors normal to $\Omega$. In the integral (\ref{X}), the functions $h=h(\cos \theta)$, $\nu = \nu(\cos \theta)$  and $M_\Omega = C \exp (\frac{\sigma(\cos \theta)}{d})$ only depend on $\theta$. Therefore, the integrals with respect to $\phi$ only concern the repeated tensor products of $\omega$. 

We first have that $\int_0^{2 \pi} \omega \, d \phi = 2 \pi \, \cos \theta \, \Omega$, 
so that 
\begin{eqnarray} 
X_1= \int_{\omega \in {\mathbb S}^2}  \partial_t \rho \, \omega \, h \, M_\Omega \, d \omega = 2 \pi \, \partial_t \rho \, \int_0^\pi \cos \theta \, h(\cos \theta) \, M_\Omega (\cos \theta) \, \sin \theta \, d \theta  \, \Omega, 
\label{X_1} 
\end{eqnarray}
and $(\mbox{Id} - \Omega \otimes \Omega) X_1 = 0$. 

Now, an easy computation shows that 
\begin{eqnarray} 
 \int_0^{2 \pi} \omega \otimes \omega \, d \phi
= \pi \sin^2 \theta \, (\mbox{Id} - \Omega \otimes \Omega) + 2 \pi \cos^2 \theta \,  \Omega \otimes \Omega .
\label{int_omega_otimes_omega}
\end{eqnarray}
We deduce that 
\begin{eqnarray} 
& & X_2 = \int_{\omega \in {\mathbb S}^2} ((\omega \otimes \omega)  \nabla_x \rho )\, h \, M_\Omega \, d \omega = \nonumber \\
& & \hspace{2cm} = \pi \int_0^\pi \sin^2 \theta \, h \, M_\Omega \, \sin \theta \, d \theta \, \, (\mbox{Id} - \Omega \otimes \Omega) \nabla_x \rho + \nonumber \\
& & \hspace{3cm} + \, 2 \pi \int_0^\pi \cos^2 \theta \, h \, M_\Omega \, \sin \theta \, d \theta \, \, (\Omega \cdot \nabla_x \rho) \,  \Omega , 
\label{X_2} 
\end{eqnarray}
which leads to:
\begin{eqnarray} 
& & \hspace{-0.5cm} (\mbox{Id} - \Omega \otimes \Omega) X_2 =
 \pi \int_0^\pi \sin^2 \theta \, h \, M_\Omega \, \sin \theta \, d \theta \, \, (\mbox{Id} - \Omega \otimes \Omega) \nabla_x \rho  , 
\label{Proj_X_2} 
\end{eqnarray}
Using (\ref{int_omega_otimes_omega}) again, we find: 
\begin{eqnarray} 
& & X_3= d^{-1} \rho \, \int_{\omega \in {\mathbb S}^2}    ((\omega \otimes \omega) \partial_t \Omega)  \, \nu \, h \, M_\Omega \, d \omega = \nonumber \\
& & \hspace{1cm} = \pi d^{-1} \rho \, \int_0^\pi \sin^2 \theta \, \nu \, h \, M_\Omega \, \sin \theta \, d \theta \, \, (\mbox{Id} - \Omega \otimes \Omega) \partial_t \Omega + \nonumber \\
& & \hspace{2cm} + \, 2 \pi d^{-1} \rho \, \int_0^\pi \cos^2 \theta \, \nu \, h \, M_\Omega \, \sin \theta \, d \theta \, \, (\Omega \cdot \partial_t \Omega) \,  \Omega .  
\label{X_3} 
\end{eqnarray}
The second term at the r.h.s. of (\ref{X_3}) vanishes since $\partial_t \Omega$ is normal to $\Omega$ ($\Omega$ being a unit vector). For the same reason, $(\mbox{Id} - \Omega \otimes \Omega) \partial_t \Omega = \partial_t \Omega$ and we are left with: 
\begin{eqnarray} 
& & \hspace{-1cm} (\mbox{Id} - \Omega \otimes \Omega) X_3=   \pi d^{-1} \rho \, \int_0^\pi \sin^2 \theta \, \nu \, h \, M_\Omega \, \sin \theta \, d \theta \, \, \partial_t \Omega .
\label{Proj_X_3} 
\end{eqnarray}

We now need to compute the integral with respect to $\phi$ of the third tensor power of $\omega$. After some computations, we are left with 
\begin{eqnarray} 
& &  \int_0^{2 \pi} \omega \otimes \omega \otimes \omega \, d \phi
= \pi \sin^2 \theta \cos \theta \, ((\mbox{Id} - \Omega \otimes \Omega) \otimes \Omega + \Omega \otimes (\mbox{Id} - \Omega \otimes \Omega) + \nonumber \\
& & \hspace{6cm} + [(\mbox{Id} - \Omega \otimes \Omega) \otimes \Omega \otimes (\mbox{Id} - \Omega \otimes \Omega)]_{:24}) \nonumber \\
& & \hspace{3.5cm}  + 2 \pi \cos^3 \theta \, \, \Omega \otimes \Omega \otimes \Omega
 ,
\label{int_omega_otimes_3}
\end{eqnarray}
where the index '$:24$' indicates contraction of the indices $2$ and $4$. In other words, the tensor element  $(\int_0^{2 \pi} \omega \otimes \omega \otimes \omega \, d \phi)_{ijk}$ equals $\pi \sin^2 \theta \cos \theta$ when $(ij,k)$ equals any of the triples $(1,1,3)$, $(2,2,3)$, $(3,1,1)$, $(3,2,2)$, $(1,3,1)$, $(2,3,2)$,  equals $2 \pi \cos^3 \theta$ when $(ij,k)= (3,3,3)$ and is equal to $0$ otherwise. 
Using Einstein's summation convention, the following formula follows: 
\begin{eqnarray} 
& &  (\int_0^{2 \pi} \omega \otimes \omega \otimes \omega \, d \phi) \nabla_x \Omega
= \left(\int_0^{2 \pi} \omega \otimes \omega \otimes \omega \, d \phi \right)_{ijk} \partial_{x_j} \Omega_{k} = 
\nonumber \\
& & = \pi \sin^2 \theta \cos \theta \, ((\mbox{Id} - \Omega \otimes \Omega)_{ij} \Omega_k \partial_{x_j} \Omega_{k} + \Omega_i (\mbox{Id} - \Omega \otimes \Omega)_{jk} \partial_{x_j} \Omega_{k}  + \nonumber \\
& & \hspace{6cm} + (\mbox{Id} - \Omega \otimes \Omega)_{ik} \Omega_j \partial_{x_j} \Omega_{k}) \nonumber \\
& & \hspace{3.5cm}  + 2 \pi \cos^3 \theta \, \, \Omega_i \Omega_j \Omega_k \partial_{x_j} \Omega_{k}
 ,
\label{int_times_nabla_Omega}
\end{eqnarray}
But since $\Omega$ is a unit vector, $\Omega_k \partial_{x_j} \Omega_{k} = \frac{1}{2} \partial_{x_j} (|\Omega|^2) = 0$ and the first and fourth terms in the sum vanish. The expression simplifies into: 
\begin{eqnarray} 
& &  (\int_0^{2 \pi} \omega \otimes \omega \otimes \omega \, d \phi) \nabla_x \Omega
=  \pi \sin^2 \theta \cos \theta \, ((\mbox{Id} - \Omega \otimes \Omega) : (\nabla_x \Omega)) \, \Omega  + \nonumber \\
& & \hspace{6cm} + \pi \sin^2 \theta \cos \theta \, (\mbox{Id} - \Omega \otimes \Omega) ((\Omega \cdot \nabla) \Omega) ,
\label{int_times_nabla_Omega_2}
\end{eqnarray}
The first term is parallel to $\Omega$. Besides, since $\Omega$ is a unit vector, $(\Omega \cdot \nabla) \Omega$ is normal to $\Omega$. So, we finally get 
\begin{eqnarray} 
& &  (\mbox{Id} - \Omega \otimes \Omega) ((\int_0^{2 \pi} \omega \otimes \omega \otimes \omega \, d \phi) \nabla_x \Omega)
=    \pi \sin^2 \theta \cos \theta \, (\Omega \cdot \nabla) \Omega ,
\label{int_times_nabla_Omega_3}
\end{eqnarray}
This leads to the following formula for $X_4$:
\begin{eqnarray} 
(\mbox{Id} - \Omega \otimes \Omega)  X_4 &=& d^{-1} \rho \, (\mbox{Id} - \Omega \otimes \Omega) \left( \int_{\omega \in {\mathbb S}^2}   (\omega \otimes \omega\otimes \omega ) ( \nabla_x \Omega)\, \nu \, h \, M_\Omega \, d \omega \right) \nonumber \\
&=& \pi d^{-1} \rho \int_0^\pi \sin^2 \theta \cos \theta \, \nu \, h \, M_\Omega \, \sin \theta \, d \theta \, \,  (\Omega \cdot \nabla) \Omega
\label{X_4} 
\end{eqnarray}

Now, we insert the expressions of $X_1$ to $X_4$ into (\ref{(Id-Omega)_X}). Using notation (\ref{brackets}),  we finally find the evolution equation for $\Omega$: 
\begin{eqnarray} 
& & \hspace{-1cm} d^{-1} \rho \, \langle  \sin^2 \theta \, \nu \, h \rangle_M \, \, \partial_t \Omega + 
d^{-1} \rho \langle \sin^2 \theta \cos \theta \, \nu \, h \rangle_M \,   (\Omega \cdot \nabla) \Omega + \nonumber \\
& & \hspace{5cm} + \langle \sin^2 \theta \, h \rangle_M \,  (\mbox{Id} - \Omega \otimes \Omega) \nabla_x \rho = 0.
\label{Eq_Omega} 
\end{eqnarray}

By the maximum principle, the function $h$ is non-positive. Therefore, we can define similar averages as (\ref{brackets}), substituting $M_\Omega$ with $\sin^2 \theta \, \nu \, h \, M_\Omega$ and we denote such averages as $\langle g \rangle_{(\sin^2 \theta) \nu  h  M}$. With such a notation, (\ref{Eq_Omega}) becomes:
\begin{eqnarray} 
& & \hspace{-1cm}  \rho \, \left( \partial_t \Omega + c_2 (\Omega \cdot \nabla) \Omega \right) +  \lambda  \,  (\mbox{Id} - \Omega \otimes \Omega) \nabla_x \rho = 0,
\label{Eq_Omega_2} 
\end{eqnarray}
with 
\begin{eqnarray} 
& & \hspace{-1cm}  c_2 = \langle \cos \theta \rangle_{(\sin^2 \theta) \nu  h  M} \, , \quad \lambda = d \left\langle \frac{1}{\nu} \right\rangle_{(\sin^2 \theta) \nu  h  M}
\label{c_2_mu} 
\end{eqnarray}
Collecting the mass and momentum eqs (\ref{mass_eq}) and (\ref{Eq_Omega_2}), we find the final macroscopic model of the Couzin-Vicsek algorithm: 
\begin{eqnarray} 
& & \partial_t \rho + \nabla_x \cdot (c_1 \rho \Omega)  = 0.
\label{mass_eq_2} \\
& &  \rho \, \left( \partial_t \Omega + c_2 (\Omega \cdot \nabla) \Omega \right) +  \lambda  \,  (\mbox{Id} - \Omega \otimes \Omega) \nabla_x \rho = 0,
\label{Omega_eq_2} 
\end{eqnarray}
with the coefficients $c_1$, $c_2$ and $\lambda$ given by (\ref{c_1}) and (\ref{c_2_mu}). This ends the proof of theorem \ref{theo_limit}.

\subsection{Hyperbolicity}
\label{subsec_hyper}

The detailed study (both theoretical and numerical) of the properties of the continuum model (\ref{mass_0}), (\ref{Omega_0}), will be the subject of future work. As a preliminary step, we look at the hyperbolicity of the model. 

First, thanks to a temporal rescaling, $t = t'/c_1$, we can replace $c_1$ by $1$, $c_2$ by $c:= c_2/c_1$ and $\lambda$ by $\lambda' =  \lambda/c_1$. We will omit the primes for simplicity. Then, the system reads: 
\begin{eqnarray} 
& & \partial_t \rho + \nabla_x \cdot (\rho \Omega)  = 0.
\label{mass_eq_3} \\
& &  \rho \, \left( \partial_t \Omega + c (\Omega \cdot \nabla) \Omega \right) +  \lambda  \,  (\mbox{Id} - \Omega \otimes \Omega) \nabla_x \rho = 0,
\label{Omega_eq_3} 
\end{eqnarray}
This rescaling amounts to saying that the magnitude of the velocity of the individual particles is equal to $1/c_1$ in the chosen system of units. 

We choose an arbitrary fixed cartesian coordinate system $(\Omega_1,\Omega_2,\Omega_3)$ and use spherical coordinates $(\theta,\phi)$ in this system (see section \ref{sec_Mean_Field}).  Then, $\Omega =(\sin \theta  \cos \phi , \, \sin \theta  \sin \phi, \, \cos \theta)$. A simple algebra shows that $(\rho,\theta, \phi)$ satisfy the system
\begin{eqnarray} 
& & \partial_t \rho + \partial_x  (\rho \sin \theta \cos \phi) + \partial_y  (\rho \sin \theta \sin \phi) + \partial_z  (\rho \cos \theta)  = 0.
\label{mass_eq_4} \\
& &  \partial_t \theta + c ( \sin \theta \cos \phi \, \partial_x  \theta + \sin \theta \sin \phi \, \partial_y  \theta + \cos \theta \partial_z \theta ) + \nonumber \\
& & \hspace{2cm} +  \lambda  \,  ( \cos \theta \cos \phi \, \partial_x \ln \rho  + \cos \theta \sin \phi \, \partial_y \ln  \rho  - \sin \theta \, \partial_z \ln \rho) = 0.
\label{theta_eq} \\
& &  \partial_t \phi + c ( \sin \theta \cos \phi \, \partial_x  \phi + \sin \theta \sin \phi \, \partial_y  \phi + \cos \theta \partial_z \phi ) + \nonumber \\
& & \hspace{2cm} +  \lambda  \,  ( - \sin \theta \sin \phi \, \partial_x \ln \rho  + \sin \theta \cos \phi \, \partial_y \ln  \rho) = 0.
\label{phi_eq}
\end{eqnarray}
Supposing that $\rho,\theta,\phi$ are independent of $x$ and $y$ amounts to looking at waves which propagate in the $z$ direction at a solid angle $(\theta,\phi)$ with the velocity director $\Omega$. In this geometry, the system reads:
\begin{eqnarray} 
& & \partial_t \rho + \cos \theta \, \partial_z  \rho - \rho \sin \theta \,  \partial_z \theta   = 0.
\label{mass_eq_5} \\
& &  \partial_t \theta + c  \cos \theta \, \partial_z  \theta -  \lambda  \,   \sin \theta \, \partial_z \ln \rho = 0.
\label{theta_eq_2} \\
& &  \partial_t \phi + c  \cos \theta \, \partial_z  \phi = 0.
\label{phi_eq_2}
\end{eqnarray}
This is a first order system of the form 
\begin{eqnarray} 
& & \hspace{-1cm}  \left( \begin{array}{c} \partial_t  \rho \\ \partial_t \theta \\ \partial_t \phi \end{array} \right) + A(\rho, \theta, \phi)  \left( \begin{array}{c} \partial_z \rho \\ \partial_z \theta \\ \partial_z \phi \end{array} \right)   = 0, \label{System} 
\end{eqnarray}
with 
\begin{eqnarray} 
A(\rho, \theta, \phi) = \left( \begin{array}{ccc} \cos \theta & - \rho \sin \theta & 0 \\ - \frac{\lambda \sin \theta}{\rho} & c \cos \theta  & 0 \\
0 & 0 & c \cos \theta \end{array} \right), 
\label{Matrix} 
\end{eqnarray}

The eigenvalues $\gamma_\pm$ and $\gamma_0$ of the matrix $A(\rho, \theta, \phi)$ are readily computed  and are given by
\begin{eqnarray} 
& & \gamma_0 = c \cos \theta , \quad 
\gamma_\pm = \frac{1}{2} \left[ 
(c+1) \cos \theta \pm \left( (c-1)^2 \cos^2 \theta + 4 \lambda \sin^2 \theta
\right)^{1/2} 
\right]. 
\label{eigenvalues} 
\end{eqnarray}

Two special cases are noteworthy. The case $\theta = 0$ (modulo $\pi$) corresponds to waves which propagate parallel to the velocity director. In this case, two eigenvalues are equal: $\gamma_0 = \gamma_+ = c$ and $\gamma_- = 1$. The eigenvectors corresponding to these three eigenvalues are respectively the density $\rho$, and the angles $\theta$ and $\phi$. So far, the relative magnitude of $c$ and $1$ are not known. But, whatever the situation ($c$ bigger or smaller or even equal to $1$), the matrix is diagonalizable and therefore the system is hyperbolic.  

The case $\theta = \pi /2$ (modulo $\pi$) corresponds to waves propagating normally to the velocity director. In this case, $\gamma_\pm = \pm 2 \sqrt \lambda$ are opposite and $\gamma_0 = 0$. The system for $(\rho,\theta)$ reduces to a special form of the nonlinear wave equation.  The sound speed which propagates in the medium due to the interactions between the particles has magnitude equal to $2 \sqrt \lambda$. 

If $\theta$ has an arbitrary value, then, a combination of these two phenomena occurs. For the two waves associated with $\gamma_\pm$, there is a net drift at velocity $(c+1)\cos \theta$ and two sound waves with velocities $\left( (c-1)^2 \cos^2 \theta \right.$ $ \left.  + 4 \lambda \sin^2 \theta \right)^{1/2}$. However, the speed of  the wave associated with $\gamma_0$, is not equal to the drift of the two sound waves. A disymmetry appears which is not present in the usual gas dynamics equations. The resolution of the Riemann problem is left to future work.

\setcounter{equation}{0}
\section{Conclusion}

In this paper,  we have studied the large-scale dynamics of the Couzin-Vicsek algorithm. For that purpose, we have rephrased the dynamics as a time-continuous one and have formulated it in terms of a kinetic Fokker-Planck equation. Then, a hydrodynamic scaling of this kinetic equation is introduced with small parameter $\varepsilon$ and the limit when $\varepsilon \to 0$ is considered. We show that the macroscopic dynamics takes place on a three dimensional manifold consisting of the density and director of the mean-velocity. Using a new concept of generalized collision invariant, we are able to derive formally the set of equations satisfied by the parameters and we prove that the resulting system is hyperbolic. 

Possible future directions involve the investigation of a limited range of vision in the backwards direction, the computation of the order $\varepsilon$ diffusive corrections, the incorporation of more non-locality effects in the asymptotics and finally, the accounting of the other types of interactions, being of repulsive or attractive type.


\end{document}